\documentstyle[12pt]{article}

\begin{document}
\title{Influence of medium correction of nucleon nucleon cross
section on the fragmentation and nucleon emission}
\author{\small Yong-Zhong Xing$^{1,2}$, Jian-Ye Liu$^{1,2,3,4}$, Wen-Jun Guo$^{3}$ \\}
\date{}
\maketitle
\begin{center}
$^{1}${\small Institute for the theory of modern physics, Tianshui Normal University, Gansu,
Tianshui 741000, P. R. China}\\
$^{2}${\small Center of Theoretical Nuclear Physics, National
Laboratory of Heavy Ion Accelerator}\\
{\small Lanzhou 730000, P. R. China}\\
$^{3}${\small Institute of Modern Physics, Chinese Academy of
Sciences, P.O.Box 31}\\
{\small Lanzhou 730000, P. R. China}\\
$^{4}${\small CCAST(Word Lab.),P.O.Box 8730,Beijing 100080}\\
{\small Lanzhou 730000, P. R. China}\\
\end{center}
\baselineskip 0.3in.
\begin{center}{\bf Abstract}\end{center}
\hskip 0.2in The influence of medium correction from an isospin
dependent nucleon nucleon cross section on the fragmentation and
nucleon emission in the intermediate energy heavy ion collisions
was studied by using an isospin dependent quantum molecular
dynamical model (IQMD). We found that the medium correction
enhances the dependence of multiplicity of intermediate mass
fragment $N_{imf}$ and the number of nucleon emission $N_{n}$ on
the isospin effect of the nucleon nucleon cross section,while the
momentum dependent interaction (MDI) produces also an important
role for enhancing the influence of the medium correction on the
isospin dependence of two-body collision in the fragmentation and
nucleon emission processes. After considering the medium
correction and the role of momentum dependent interaction the
increase for the dependence of $N_{imf}$ and $N_{n}$ on the
isospin effect of two-body collision is favorable to learn the
information about the isospin dependent
nucleon nucleon cross section.
{\bf PACCS}: 25.70.Pq, 02.70.Ns, 24.10.Lx
\baselineskip 0.3in.
\section{Introduction}
   The isospin physics in heavy-ion collision (HIC) at intermediate
energies has been an important topic in recent years[1,3,18].
These studies is not only important for understanding the
collision mechanism and nuclear structure but also for getting the
knowledge about the isospin asymmetric nuclear matter equation of
state(EOS) and isospin dependent nucleon nucleon cross section.
Recently, some observables  were found to be the good probes for
extracting the information of an isospin asymmetric nuclear matter
EOS and the in-medium nucleon nucleon cross section at
intermediate energies [4-29] because they are sensitive only to
one of the isospin dependent mean field and  the isospin dependent
in-medium nucleon nucleon cross section. Bao-an Li[22] has pointed
out, for instance,that the proton neutron differential collective
flow and proton ellipse flow can be used to probe the isospin
asymmetric nuclear matter equation of state. He also found
recently that the isospin asymmetry of the high density nuclear
matter formed in high energy heavy-ion collision is uniquely
determined by the high density behavior of the nuclear symmetry
energy[23]. Our studies in last few years indicated that the
nuclear stopping, the number of nucleon emission and the
multiplicity of intermediate mass fragments in HIC at intermediate
energies  can be used to probe the isospin-dependent in-medium
nucleon nucleon cross section[24,25]. In this work we investigate
further the influence of the medium correction of the isospin
dependent nucleon nucleon cross section on the fragmentation and
the nucleon emission in the heavy ion collisions at intermediate
energies by using an isospin dependent quantum molecular dynamical
model. We found that the multiplicity of intermediate mass
fragments $N_{imf}$ and the number of nucleon emission $N_{n}$ for
the medium effect $\alpha = -0.2$( see in Eq.(7)) are always less
than those for $\alpha = 0.0$ ( free nucleon nucleon collision) in
the intermediate energy heavy ion collisions. In particular,the
differences between $N_{imf}$'s or $N_{n}$'s from an isospin
dependent nucleon nucleon cross section and an isospin independent
one with  $\alpha = -0.2$ are always larger than those with
$\alpha = 0.0$,i.e., the medium correction of two-body collision
increases the dependence of $N_{imf}$ and $N_{n}$ on the isospin
effect of nucleon nucleon cross section, while MDI enhances also
the influence of the medium correction on the isospin effect of
two-body collision in the fragmentation and nucleon emission
processes.

{\section {IQMD model}} The quantum molecular dynamics(QMD)[30]
contains two ingredients: density dependent mean field and
in-medium nucleon nucleon cross section. To describe isospin
effects appropriately, QMD should be modified properly: the
density dependent mean field should contain correct isospin terms
including symmetry  potential and coulomb potential, the in-medium
nucleon nucleon cross sections should be different for neutron
neutron (proton proton) and neutron proton collisions, in which
the Pauli blocking should be counted by distinguishing neutrons
and protons. In addition, the initial condition of the ground
state of two colliding nuclei should also contain isospin
information.\\

 Considering the above ingredients, we have made important modifications in  QMD to
obtain an isospin dependent quantum molecular
dynamics(IQMD)$[1,3]$. The initial density distributions of the
colliding nuclei in  IQMD are obtained from the calculations of
the Skyrme-Hatree-Fock with parameter set SKM$^{*}$ $[39]$.  The
initial code of IQMD was used to determine the ground state
properties of the colliding nuclei, such as the binding energies
and RMS radii,which agree with the experimental data for obtaining
the parameters of interaction potential as an input data for the
collision dynamics calculations by using
the code of IQMD.\\\
 The interaction potential is
\begin{equation}
U(\rho)=U^{Sky}+U^{c}+U^{sym}+U^{Yuk}+U^{MDI}+U^{Pauli}
\end{equation}
,where $U^{c}$ is Coulomb potential.

The density dependent Skyrme potential $U^{Sky}$, the Yukawa potential $U^{Yuk}$,
the momentum dependent interaction $U^{MDI}$ and the Pauli potential $U^{Pauli}$
$[30,37]$ are given by the following equations, respectively
\begin{equation}
U^{Sky}=\alpha (\frac \rho {\rho _0})+\beta (\frac \rho {\rho _0})^\gamma  ,
\end{equation}
\begin{equation}
U^{Yuk}=t_3exp(\frac{\left|\overrightarrow{r_1}-
\overrightarrow{r_2}\right|}{m})/\frac{\left|\overrightarrow{r_1}-
\overrightarrow{r_2}\right|}{m},
\end{equation}
\begin{equation}
U^{MDI}=t_4ln^2[t_5(\overrightarrow{p_1}-\overrightarrow{p_2})^2+1]\frac
\rho {\rho _0},
\end{equation}
and
\begin{equation}
U^{Pauli}=V_p\{\frac \hbar {p_0q_0})^3exp(-\frac{(\overrightarrow{r_i}-%
\overrightarrow{r_j})^2}{2q_0^2}-\frac{(\overrightarrow{p_i}-\overrightarrow{%
p_j})^2}{2p_0^2}\}\delta _{p_ip_j},
\end{equation}

with
 $$\delta_{p_{i}p_{j}}=\left\{ \begin{array}{ll}
              1 & \mbox{for neutron-neutron or proton-proton}\\
              0 & \mbox{for neutron-proton}.
             \end{array}
           \right .$$

In this work we used following different symmetry potentials[1,3]:
$$\begin{array}{ll}
U^{sym}_{1}=\pm 2e_{a}u\delta\\
U^{sym}_{2}=\pm
2e_{a}u^{2}\delta+e_{a}u^{2}\delta^{2}\\
U^{sym}_{0}=0.0
\end{array}                                          \eqno  (6)$$
Here $e_{a}$ is the strength of symmetry potential taking the
value of 16MeV and $U^{sym}_{0}$=0.0 indicates the case without
any symmetry potential. $u= \frac {\rho}{\rho_0},\delta$ is the
relative neutron excess $\delta =\frac{\rho _n-\rho _p} {\rho
_n+\rho _p}=\frac{\rho _n-\rho _p}\rho $. Here $\rho $,$\rho
_{_0}$,$\rho _n$ and $\rho _p$ are total, normal, neutron and
proton densities, respectively. In the first terms on the right
side of Eq.(6) the upper + means repulsive for neutrons and the
lower - is attractive for protons. The parameters of interaction
potentials are in table 1.0

 Table 1. The parameters of the interaction potential
\begin{center}
\scriptsize
\begin{tabular}{|c|c|c|c|c|c|c|c|c|c|c|} \hline
\small
 &$\alpha$ & $\beta$ &$\gamma$&$t_{3}$&m&$t_{4}$&$t_{5}$&$V_{p}$&$p_{0}$&$q_{0}$\\ \hline
 &(MeV)&(MeV)&&(MeV)&(fm)&(MeV)&($MeV^{-2}$)&(MeV)&(MeV/c)&(fm)\\\hline
MDI&-390.1&320.3&1.14&7.5&0.8&1.57&$5\times10^{-4}$&30&400&5.64\\\hline
NOMDI&-356&303&1.1667&7.5&0.8&0.0&$0\times10^{-4}$&30&400&5.64\\\hline
\end{tabular}\\
\end{center}
The NOMDI in table 1 means without MDI. The influence of medium
correction on the nucleon nucleon cross section is an important
topic in HIC at intermediate energies[31,32].  D. Klakow et al
proposed that the in-medium nucleon nucleon cross section should
be a function of the nucleon distribution density as follows [33]
$$\sigma_{NN}=(1+\alpha\frac{\rho}{\rho_{0}})\sigma^{free}_{NN}.  \eqno (7)$$
The parameter $\alpha =-0.2$ has been found to reproduce the flow
data[34,35]. The free neutron proton cross section is about a
factor of 3 times larger than the free proton proton or the free
neutron neutron one below 400 MeV, which contributes the main
isospin effect from nucleon-nucleon collisions at intermediate
heavy ion collisions. In fact, the ratio of the neutron proton
cross section to proton proton(or neutron neutron) cross section
in the medium, $\sigma_{np}/\sigma_{pp}$, depends sensitively on
the evolution of the nuclear density distribution and beam energy.
We used equation (7) to take into account the medium effects, in
which the neutron proton cross section is always larger than the
neutron neutron or proton proton cross section in the medium at
the beam energies in this paper. Here $\sigma^{free}_{NN}$ is the
experimental nucleon nucleon cross section[36].
$$\sigma^{free}_{np}=\left\{ \begin{array}{llll}
                           381.0(mb) & \mbox{E $\leq$ 25(MeV)}\\
 \frac{5067.4}{E^{2}}+\frac{9069.2}{E}+6.9466(mb) & \mbox{25 $<$ E  $\leq$ 40 (MeV)}\\
 \frac{239380.0}{E^{2}}+\frac{1802.0}{E}+27.14(mb) & \mbox{40 $< $ E $\leq$ 310 (MeV)}\\
34.5(mb)                                          & \mbox{310 $<$ E $\leq$ 800(MeV)}
 \end{array}\right.
\eqno  (8)$$
$$\sigma^{free}_{nn}=\sigma^{free}_{pp}=
\left\{ \begin{array}{llll}
                           80.6(mb) & \mbox{E$\leq$  25(MeV)}\\
 -\frac{1174.8}{E^{2}}+\frac{3088.5}{E}+5.3107(mb) & \mbox{25$<$  E $\leq$ 40 (MeV)}\\
 \frac{93074.0}{E^{2}}+\frac{11.148}{E}+22.429(mb) & \mbox{40 $< $ E$\leq$  310 (MeV)}\\
 \frac{887.37}{E^{2}}+0.05331E+3.5475(mb)& \mbox{310$ < $  E$\leq$  800(MeV)}
 \end{array}\right.
\eqno  (9)$$
\par
We constructed the clusters by means of a modified coalescence model $[37]$, in which particle
relative momentum is smaller than $p_{0}$=300MeV/c and relative distance is smaller than
$R_{0}$= 3.5 fm. The restructured aggregation model$[38]$ has been applied to avoid the
nonphysical clusters after constructing the clusters, until there were not any nonphysical
clusters to be produced.
{\section{Results and Discussions}}
The isospin effect of the in-medium nucleon nucleon
cross section on the observables is defined by the difference
between the observables for an isospin dependent nucleon nucleon
cross section  $\sigma^{iso}$ and for an isospin independent
one $\sigma^{noiso}$ in the medium. Here
$\sigma^{iso}$ is defined as $\sigma_{np} \geq
\sigma_{nn}$=$\sigma_{pp}$ and $\sigma^{noiso}$ means
$\sigma_{np}$ = $\sigma_{nn}$ = $\sigma_{pp}$, where
$\sigma_{np}$, $\sigma_{nn}$ and $\sigma_{pp}$ are the neutron
proton, neutron neutron and proton proton cross sections in medium,
respectively.

{\subsection{Influence of medium correction of two-body
(collision) on the $N_{n}$ and $N_{imf}$}}
 In order to study the influence from the medium correction of the
isospin dependent nucleon nucleon cross section on the isospin
effects of the fragmentation and the nucleon emission, we
investigated the number of nucleon emission $N_{n}$ as a function
of the beam energy at impact parameter b= 4.0 fm for the mass
symmetry system $^{76}K_{r}+^{76}K_{r}$ (top panels) and mass
asymmetry system $^{112}S_{n}+^{40}C_{a}$ (bottom panels). Two
colliding systems are the same system mass $A_{t}+A_{p}=152$,
where $A_{t}$ and $A_{p}$ are projectile mass and target mass
respectively. The different symmetry potentials $U_1^{sym}$ ,
$U_2^{sym}$ and $U_0^{sym}$ as well as different kinds of nucleon
nucleon cross sections, i.e., the isospin dependent in-medium
nucleon nucleon cross section $\sigma^{iso}$ and the isospin
independent one $\sigma^{noiso}$ are used here. Namely, there are
six cases : $U_0^{sym}$+$\sigma^{iso}$ ,$U_1^{sym}$+$\sigma^{iso}$
,$U_2^{sym}$+$\sigma^{iso}$ , $U_1^{sym}$+$\sigma^{noiso}$ and
$U_2^{sym}$+$\sigma^{noiso}$ with
 $\alpha =-0.2$ (left panels) and $\alpha = 0.0$ (right panels) in Fig.1.
The detail explanations about line symbols in the Fig.1 are in
figure.
  It is clear to see that all of lines with filled symbols are larger
than those with open symbols, i.e., all of $N_{n}'s$ with
$\sigma^{iso}$ are larger than those with $\sigma^{noiso}$ because
the collision number is larger for $\sigma^{iso}$ than that for
$\sigma^{noiso}$. We also found that the gaps between lines with
the filled symbols and with the open symbols are larger but the
variations among lines in each group are smaller. Because the
large gaps come from the isospin effect of two-body collision and
small variations are produced from the symmetry potential, i. e.,
$N_{n}$ depends sensitively on the isospin effect of in-medium
nucleon nucleon cross section and weakly on the symmetry
potential. In particular, the gaps between two group lines with
$\alpha =-0.2$  are larger than those with $\alpha = 0.0$, i.e.,
the medium correction of two-body collision enhances the
dependence of $N_{n}$ on isospin effect of two-body collision.

In order to investigate quantitatively the influence of the medium
correction of two-body collision on the nucleon emission, it is to
define
$$\Delta N_{n}(\Delta \sigma , \alpha)= N_{n}(\sigma^{iso}, \alpha)
-N_{n}(\sigma^{noiso}, \alpha).             \eqno(10)$$ Fig.2
shows the time evolution of $\Delta N_{n}(\Delta \sigma, \alpha
=-0.2$) (solid line) and $\Delta N_{n}(\Delta \sigma, \alpha =
0.0$) (dashed line)  at impact parameter $b=4.0fm$ for $^{76}$Kr +
$^{76}$Kr at E=100 MeV/nucleon (top panels) and $^{112}$Sn
+$^{40}$Ca at E=200 MeV/nucleon (bottom panels). In this case,
there are about the same center of mass energy per nucleon for two
colliding systems. The symmetry potentials are $U^{sym}_{1}$ (left
panels) and $U^{sym}_{2}$ (right panels). It is clear to see that
all of solid lines are higher than dashed lines,i.e.,the medium
correction of nucleon nucleon cross section increases sensitively
the dependence of $N_{n}$ on the isospin effect of two-body
collision as above mentioned .

To investigate the evolution of above dependence with increasing
beam energy for two different colliding systems, the variation
$\Delta N_{n}( \Delta \sigma, \Delta \alpha)$ as a function of
beam energy at impact parameter of 4.0 fm for symmetry potential
$U_{1}^{sym}$ is given in Fig.3. Where  $\Delta N_{n}( \Delta
\sigma, \Delta \alpha)$ is defined as
$$\Delta N_{n}( \Delta \sigma, \Delta \alpha)
=\Delta N_{n}( \Delta \sigma, \alpha =-0.2) -\Delta N_{n}( \Delta
\sigma,  \alpha =0.0)                 \eqno (11).$$ In which
$\Delta N_{n}( \Delta \sigma,\alpha)$ is taken from Eg.(10). From
Fig.3 we can see that all of $\Delta N_{n}( \Delta \sigma, \Delta
\alpha)$ are larger than zero, i.e., the medium correction of the
nucleon nucleon cross section increases the dependence of $N_{n}$
on the isospin effect of two-body collision in the beam energy
region from 50 MeV/nucleon to 300 MeV/nucleon. we also find that
$\Delta N_{n}( \Delta \sigma, \Delta \alpha)$'s for mass symmetry
system are always larger than those for mass asymmetry system due
to more collision number for the mass symmetry system with the
same system mass.

In order to investigate the contributions from all of impact
parameters to $N_{n}$ , Fig.4 shows the impact parameter average
values of $<N_{n}>_{b}$ (from at equilibrium time $ 200 fm/c $) as
a function of the beam energy for above two colliding systems with
the same incident channel conditions and line symbols as in Fig.1.
The same conclusion as  mentioned in Fig.1 is also obtained here,
i.e., the medium correction of nucleon nucleon cross section
enhances the dependence of $<N_{n}>_{b}$ on the isospin effect of
two-body collision.

 Fig.5 shows the impact parameter average values of the multiplicity of
intermediate mass fragment, $<N_{imf}>_{b}$(at equilibrium time $
200 fm/c $), as a function of the beam energy for the same
incident channel conditions and line symbols as in Fig.4. Where
the charge number of intermediate mass fragment is taken from 3 to
13. From Fig.5 we got the same conclusions as $<N_{n}>_{b}$,
namely, $<N_{imf}>_{b}$ depends sensitively on the isospin effect
of in-medium nucleon nucleon cross section and weakly on the
symmetry potential. In particular, the gaps between two group
lines with $\alpha=-0.2$  are larger than those with $\alpha=0.0$,
which indicates that the medium correction enhances also the
dependence of $<N_{imf}>_{b}$ on the isospin effect of the
two-body collision.
 {\subsection {Important role of MDI on
$N_{imf}$ and $N_{n}$ in the medium corrections of two-body
collision}}

We also found an important role of the MDI on $N_{imf}$ and
$N_{n}$ in the medium correction of two-body collision.
 Fig.6 shows the time evolution of
$N_{imf}$ for the reaction $^{76}Kr+^{76}Kr$ with symmetry
potentials $U_1^{sym}$ at beam energy of 100 MeV/nucleon and
impact parameter of $4.0fm$. They are four cases: (1) $\alpha
=-0.2$ +$\sigma^{iso}$ ( solid line),(2) $\alpha = 0.0$
+$\sigma^{iso}$ ( dashed line),(3) $\alpha =-0.2$
+$\sigma^{noiso}$ ( dot line ) and (4) $\alpha = 0.0$
+$\sigma^{noiso}$ ( dot-dashed line) with MDI in the left window
and  NOMDI in the right window. It is clear to see that the gap
between the lines with MDI  is larger than corresponding gap
between lines with NOMDI  in the medium($\alpha =-0.2$), i.e., MDI
increases the isospin effect of two-body collision on the
$N_{imf}$ in the medium because above gaps are produced from the
isospin effect of nucleon-nucleon cross section in the medium.

For the $<N_{n}>_{b}$, we have gotten the same conclusion as
$<N_{imf}>_{b}$ in Fig.6.

{\subsection{Explanations for the medium correction of two-body
collision and the role of MDI on the $N_{imf}$ and $N_{n}$.}}
 Why does the medium correction of nucleon nucleon cross section and MDI enhance the
dependences of multiplicity of intermediate mass fragments
$N_{imf}$ and the number of nucleon emission $N_{n}$ on the
isospin effect of two-body collision? Physically there are three
mechanisms at work here. (1) The average momentum of a particle in
medium is higher in a heavy ion collision than in cold nuclear
matter at the same density. (2) MDI induces the transporting
momentum more effectively from one part of the system to another,
in which particles also move with a higher velocity in the medium
than in free space for a given momentum . (3)As well know that the
isospin dependent in-medium nucleon nucleon cross section is a
sensitive function of the nuclear density distribution and beam
energy as shown in Eg.(7). Fig.7 shows the time evolution of the
ratio of nuclear density to normal one, $\frac{\rho} {\rho_{0}}$,
for four cases :  they are $\rho(\sigma^{iso},\alpha = -0.2)$
(solid line),$\rho(\sigma^{noiso},\alpha = -0.2)$ (dotted
line),$\rho(\sigma^{iso},\alpha = 0.0)$ (dashed line) and
$\rho(\sigma^{noiso},\alpha = 0.0)$ (dot-dashed line) for the
reaction $^{76}$Kr + $^{76}$Kr with symmetry potential $U_1^{sym}$
at E= 150 MeV/nucleon and b= 4.0fm. From the values of peak for
$\frac{\rho} {\rho_{0}}$ in the insert in Fig.7 it is clear to see
that  $\rho(\sigma^{iso},\alpha = -0.2)$ (solid line) is larger
than $\rho(\sigma^{noiso},\alpha = -0.2)$ (dot line) and
$\rho(\sigma^{iso},\alpha = 0.0)$ (dashed line)is larger than
$\rho(\sigma^{noiso},\alpha = 0.0)$ (dot-dashed line) because the
larger collision number from $\sigma^{iso}$ increases the nuclear
stopping and dissipation , which enhances the nuclear density,
compared to the case with $\sigma^{noiso}$. From Fig.7 we can also
see that $\frac{\rho} {\rho_{0}}$  decreases quickly with
increasing the time after the peak of $\frac{\rho} {\rho_{0}}$.
During above process the larger compression produces quick
expanding process of the colliding system and the small
compression induces slow expanding process ,at the same time, the
$\frac{\rho} {\rho_{0}}$ decrease quickly with expanding process
of system. But the decreasing velocity of $\frac{\rho} {\rho_{0}}$
is larger for the quick expansion system than that for the slow
expansion system, up to about after 70 fm/c, on the contrary,
$\rho(\sigma^{noiso},\alpha = -0.2)$ (dot line) is larger than
$\rho(\sigma^{iso},\alpha = -0.2)$ (solid line) and
$\rho(\sigma^{noiso},\alpha = 0.0)$ (dot-dashed line) is larger
than $\rho(\sigma^{iso},\alpha = 0.0)$ (dashed line). In
particular, the gap between two lines for $\alpha = -0.2$ is
larger than that for $\alpha = 0.0$ after about 70 fm/c. This
property is very similar to the $N_{imf}$ and $N_{n}$, which means
that the medium correction of an isospin dependent nucleon nucleon
cross section enhances also the dependence of $\rho$ on the
isospin effect of two-body collision , which induces the same
effects on $N_{imf}$ and $N_{n}$ through the nucleon nucleon cross
section as a function of the nuclear density as shown in Eg.(7).

{\section {Summary and conclusion}} \hskip 0.3in We studied the
influences of the medium correction of the isospin dependent
nucleon nucleon cross section and MDI on the fragmentation and
nucleon emission in the heavy ion collisions at intermediate beam
energies by using the IQMD. From the calculation results we can
get the following conclusions:

(1) $<N_{n}>_{b}$ and  $<N_{imf}>_{b}$ depend sensitively on the isospin effect of nucleon
 nucleon cross section and weakly on the symmetry potential.

(2) In particular, the medium correction of nucleon nucleon cross section enhances
the dependence of $<N_{n}>_{b}$ and  $<N_{imf}>_{b}$ on the isospin effect of nucleon
nucleon cross section in the intermediate beam energy region.

(3) MDI produces an important role for enhancing the isospin
effect of two-body collision  on the $<N_{n}>_{b}$ and
$<N_{imf}>_{b}$ due to the medium correction.

\vskip 1.0cm

\section{ACKNOWLEDGMENT}
We thank Prof.Bao-An Li for helpful discussions.\\\
This work is supported by the Major State Basic Research
Development Programme of China under Grant No G2000077400, the
National Nature Science Foundation Of China under Grant No.
10175080, 10004012, 19847002 and 10175082;100 person project of
Chinese Academy of Sciences and Knowledge Innovation Project of
Chinese Academy of Sciences under Grant No KJCX2-SW-N02.

\newpage

\baselineskip 0.2in
\section*{Figure captions}
\begin{description}
\item[Fig. 1] The nucleon emission number $N_{n}$ as a function of the beam energy for
systems $^{76}K_{r}+^{76}K_{r}$ and $^{112}S_{n}+^{40}C_{a}$ for
six cases ( see text).
\item[Fig. 2]
The time evolution of $\Delta N_{n}( \Delta \sigma, \alpha )$ for
the systems as the same as Fig.1 at E=100 MeV/nucleon (top panels)
and 200 MeV/nucleon (bottom panels) for two symmetry potentials
(see text).
\item[Fig. 3]$\Delta N_{n}(\Delta \sigma,\Delta \alpha)$ as a function
of the beam energy for the systems as the same as Fig.1 in two
cases (see text).
\item[Fig. 4]
The impact parameter average value of the number of nucleon
emission $<N_{n}>_{b}$ as a function of beam energy for the same
incident channel conditions and line symbols as Fig.1 (see text).
\item[Fig. 5]
The impact parameter average values of the multiplicity of
intermediate mass fragment, $<N_{imf}>_{b}$ as a function of the
beam energy for the same incident channel conditions and line
symbols as Fig.4.
\item[Fig.6]
 The time evolution of the $ N_{imf}$ with MDI
 (solid line) and NOMDI (dot line) for systems $^{76}Kr+^{76}Kr$
at E=100 MeV/nucleon and b= 4.0 fm(see text).
\item[Fig.7]
The time evolution of the ratio of nuclear density to normal density
 $\frac{\rho(\sigma^{iso},\alpha )} {\rho_{0}}$ for four cases(see text).

\end {description}


\newpage
\begin{thebibliography}{99}

\baselineskip 0.28in
\bibitem{s1} B. A. Li and W. Udo Schroder,Isospin Physics in Heavy-Ion Collision
at intermediate Energies,  Nova Science publishers. Inc(2001,New York).
\bibitem{s2} M. S. Hussein, R. A. Rego and C. A. Bertulani, Phys. Rep. 201,279(1993).
\bibitem{s3} Bao-An Li, Che-Ming Ko and W. Bauer, Int. J. Mod. Phys.
           {\bf E7},147(1998).
\bibitem{s4} R. Wada et al., Phys. Rev. Lett., {\bf 58},1829(1987).
\bibitem{s5} S. J. Tennello et al., Phys. Lett. {\bf B321}, 14(1994);
             Nucl. Phys. {\bf  A681}, 317c(2001).
\bibitem{s6} R. Pak et al., Phys. Lett. {\bf 78}, 1022(1997); ibid {\bf 78}, 1026(1997).
\bibitem{s7} G. D. Westfall, Nucl. Phys. {\bf A630}, 27c(1998); ibid {\bf A681},343c(2001).
\bibitem{s8} G. J. Kunde et al., Phys. Rev. Lett., {\bf 77}, 2897(1996).
\bibitem{s9} M. L. Miller et al., Phys. Rev. Lett., {\bf 82}, 1399(1999).
\bibitem{s10} H. Xu et al., Phys. Rev. Lett., {\bf 85}, 716(2000); M. B. Tsang et al.,
              Phys. Rev. Lett., {\bf 86}, 5023(2001).
\bibitem{s11} W. Udo Schroder, et al., NUcl. Phys. {\bf A681}, 418c(2001).
\bibitem{s12} L. G. Sobotka et al., Phys. Rev. {\bf C55}, R1272(1994).
\bibitem{s13} F. Rami et al., Phys. Rev. Lett., {\bf 84}, 1120(2000).
\bibitem{s14} M. Farine, T. Sami, B. Remaud et al., Z. Phys. {\bf 339}, 363(1991).
\bibitem{s15} H. Muller and B. D. Serot, Phys. Rev. {\bf C52}, 2072(1995).
\bibitem{s16} B. A. Li et al., Phys. Rev. Lett., {\bf 76}, 4492(1996);
              78, 1644(1997).
\bibitem{s17} G. Kortmeyer, W. Bauer and G. J. Kunde, Phys. Rev. {\bf C55} 2730(1997).
\bibitem{s18} M. Colonna et al., Phys. Lett. {\bf B428}, 1(1998);
              V. Baran et al., Nucl. Phys. {\bf A632}, 287(1998);V.Baran,M.Colonna,M.Di Toro
               et al,  Nucl. Phys. {\bf A703}, 603(2002).
\bibitem{s19} J. Pan and S. Das Gupta, Phys. Rev. {\bf C57}, 1839(1998).
\bibitem{s20} P. Chomaz and F. Gulminelli, Phys. Lett. {\bf B447}, 221(1999).
\bibitem{s21} J. Y. Liu et al., Phys. Rev. {\bf C63} 054612(2001);
              Nucl. Phys. {\bf A687}, 475(2001).
\bibitem{s22} B. A. Li, Phys. Rev. Lett., {\bf 85}, 4221(2000); {\it Phys.Rev.}{\bf C64},
054604(2001).
\bibitem{s23} Bao-An Li, Phys. Rev. Lett. {\bf 88},192701(2002).
\bibitem{s24} J. Y. Liu et al., Phys. Rev. Lett., {\bf 86}, 975(2001);
              J. Y. Liu, Y. Z. Xing, W. J. Guo, et al., Chin. Phys.
                  Lett.,19(8),1078(2002).
\bibitem{s25} J. Y. Liu, W. J. Guo, Y. Z. Xing et al.,Phys. Lett. {\bf B540},213(2002).
\bibitem{26}  J. Y. Liu, W. J. Guo, Y. Z. Xing et al., Phys. Rev. {\bf C67} 024608(2003).
\bibitem{s27} D.R.Bowman, C.M.Mader et al., Phys. Rev. {\bf C46},1834(1992).
\bibitem{s28} M.L.Miller,O.Bjarki et al., Phys.Rev.Lett. {\bf 82}, 1399(1999).
\bibitem{s29} R. Pak, Bao-An Li, W. Benenson, et al., Phys. Rev. Lett. {\bf 78},1026(1997).
\bibitem{s30} J. Aichelin, G. Peilert, A. Bohnet, et al., Phys.Rev. {\bf C37},2451(1988).
\bibitem{s31} G. Q. Li and R. Machleidt, Phys. Rev. {\bf C49} 566(1994) and references therein.
\bibitem{s32} Q. F. Li, Z. X. Li and G. J. Mao, Phys. Rev. {\bf C62}, 014606(2000);
              Q. F. Li Z. X. Li, Chin. Phys. Lett., 19(3),321(2002).
\bibitem{s33} D.Klakow, G.Welke and W.Bauer, Phys. Rev. {\bf C48},1982(1993).
\bibitem{s34} M. J. Huang et al.,  Phys.Rev.Lett., {\bf 77},3739(1996).
\bibitem{s35} G. D. Westfall et al., Phys.Rev.Lett., {\bf 71},1986(1993).
\bibitem{s36} K. Chen, Z. Fraenkel, G. Friedlander, et al., Phys. Rev. 166, 949(1968).
\bibitem{s37} G. F. Bertsch and S. D. Gupta, Phys.Rep. {\bf 160},1991(1988).
\bibitem{s38} C. Ngo, H. Ngo and S. Leray et al., Phys. Rep. {\bf A499}, 148(1989).
\bibitem{s39} P.G.Reinhard et al,in:Computational Nuclear Physics.Vol.1,Springer-Verlag,
Berlin,1991,pp.28-50.
\end{thebibliography}
\end {document}